# The Infrared Imaging Spectrograph (IRIS) for TMT: the atmospheric dispersion corrector

Andrew C. Phillips*[a], Brian J. Bauman[b], James E. Larkin[c], Anna M. Moore[d], Cynthia N. Niehaus[c], David Crampton[e], Luc Simard[e]

[a]University of California Observatories, CfAO, Univ. of California, Santa Cruz, CA, USA 95064;
[b]Lawrence Livermore National Lab, 7000 East Ave, M/S L-210, Livermore, CA USA 94550; [c]Dept. of Physics & Astronomy, Univ. of California, Los Angeles, CA, USA 90095-1547; [d]Caltech Optical Observatories, 1200 E California Blvd, M/S 11-17, Pasadena, CA, USA 91125; [e]Dominion Astrophysical Observatory, National Research Council, 5071 W Saanich Rd, Victoria, BC, Canada V9E 2E7

**ABSTRACT**

We present a conceptual design for the atmospheric dispersion corrector (ADC) for TMT's Infrared Imaging Spectrograph (IRIS). The severe requirements of this ADC are reviewed, as are limitations to observing caused by uncorrectable atmospheric effects. The requirement of residual dispersion less than 1 milliarcsecond can be met with certain glass combinations. The design decisions are discussed and the performance of the design ADC is described. Alternative options and their performance tradeoffs are also presented.

## 1. INTRODUCTION

The Infrared Imaging Spectrograph (IRIS)[1] is a first-light instrument for the Thirty-Meter Telescope (TMT). It is located behind the near-infrared facility adaptive optics system (NFIRAOS)[2], and contains both an imager and an integral-field unit (IFU) spectrograph. The spectrograph has both lenslet and image-slicer modes. It is designed to operate over the wavelength range from 0.84 to 2.4 microns.

At the fine spatial resolution of adaptive optics on the TMT, astrometric precisions of several tens of microarcseconds (μas) should be achievable. However, at these scales atmospheric dispersion across even near-IR passbands becomes problematic. This is illustrated by Figure 1, which shows the dispersion at different zenith distances of the blue edge relative to the red edge in various passbands. Across a single passband at lower elevations, differential dispersions of ~100 milliarcseconds (mas) are typical. Such dispersions will lead to significant image blur, and the integrated image position will depend on the spectral energy distribution (SED) of the objects being imaged. Therefore, dispersions of this magnitude must be largely corrected in order to reach the astrometric precision desired.

This paper describes the conceptual design for an atmospheric dispersion corrector (ADC) to be built as an integral part of IRIS. The requirements and design options are listed in Section 2, followed by an exploration of which glass pairs are suitable for the ADC, described in Section 3. In Section 4, we discuss the actual optical design of the ADC. We conclude in Section 5 with a discussion of practical limitations for ADCs operating in the adaptive optics regime on next-generation large telescopes.

*phillips@ucolick.org; phone 1 831 459-2476; fax 1 831 459-5717; www.ucolick.org

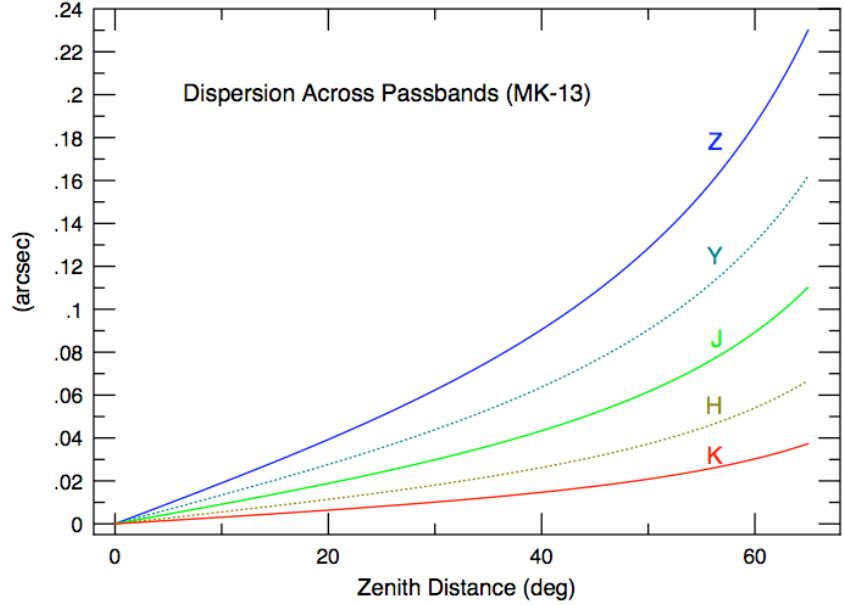

Figure 1. Atmospheric dispersion within the various passbands covered by IRIS, as a function of zenith distance. This figure shows the dispersion of the blue edge relative to the red edge of each passband. The dispersion has been calculated for the adopted site of TMT on Mauna Kea.

## 2. REQUIREMENTS AND SPECIFICATIONS

### 2.1 Requirements

The adopted requirements for the IRIS ADC are listed in Table 1. In addition to the listed requirements, the glass pairs should have high transmission, and distortion should be manageable. It was decided early on that there would be separate ADCs for the Imager, the IFU and the on-instrument wave-front sensors (OIWFS). Also, because of potential difficulties between the OIWFS (which operates with integrated light out to 1.7 μm), and IRIS (which operates in a single passband at a given time), it was decided that we would not request the removal of any IRIS aberrations by NFIRAOS.

Table 1. Adopted ADC Requirements.

| Wavelength Range | $0.84 \leq \lambda \leq 2.40$ μm in passbands Z, Y, J, H, K |
|---|---|
| Residual dispersion | ±1 mas or better within each passband |
| Zenith Distance | $1° \leq Z \leq 65°$ |
| Field of view: Imager | 15 arcsec square (r = 10.8 arcsec) |
| Field of view: IFU | 4.4 × 2.25 arcsec (r = 2.5 arcsec) |

For atmospheric dispersion, we adopted the SLALIB[3] model atmospheric refraction for Mauna Kea and a zenith distance of 65°. We found that at smaller zenith distances, the dispersion can simply be scaled by the overall refraction with negligible second-order effects.

### 2.2 Choice of ADC Design

Two principle designs are available for ADCs. Both work by producing a variable dispersion in one axis that is designed by glass choice to mimic the dispersion of the atmosphere; this dispersion is aligned along the parallactic angle in the sense that it cancels the atmospheric dispersion.

Perhaps the more traditional design is the crossed Amici prisms (often called "crossed Risley prisms"). In this design, a pair of identical counter-rotating prisms produces an optical system where the dispersion in one axis is variable, while in the perpendicular axis dispersion is internally cancelled. This design works in a collimated beam. In order to avoid tilting the optical axis of the system as the dispersion is varied, each prism is actually an Amici prism composed of a pair of glasses (one high dispersion, one low dispersion) that produces zero deviation at some chosen wavelength. The prisms are counter-rotated to produce double the dispersion of a single prism (0° with respect to each other) to no dispersion (180° with respect to each other, forming a simple plate). This kind of system produces variable distortion but is otherwise aberration free in a perfectly collimated beam.

The second design is the Linear or Longitudinal design[4], wherein two identical prisms (one rotated 180° with respect to the other) are placed in the converging beam, and the spacing of the prisms is varied. This means that each monochromatic image is displaced by differing amounts, so that they can be "stacked" into a common position. Maximum correction is when the prisms are spaced furthest apart; at zero spacing the system becomes a simple plate with no dispersion. The linear ADC can be made with simple prisms (for example, in the Keck Cassegrain ADC[5]), although in this case it displaces the optical axis of the system as the prisms separate, so zero-deviation compound prisms are usually chosen (as in, for example, the Gemini Planet Imager[6]). This design introduces a small amount of spherical aberration and some "constant" coma. To minimize the diameter of the prisms, they are usually placed near the focal plane of the telescope.

The performance of these designs is quite similar. In choosing one design over another, the deciding factors are primarily questions of packaging and whether collimated space is available. The small aberrations introduced by the Linear ADC are usually tolerable for seeing-limited instruments, and can be removed in adaptive optics systems. For IRIS, we find that (a) the lack of space near the focal plane; (b) the availability of collimated space; and (c) the unavailability of the adaptive optics system to correct instrumental aberrations, all drive us to adopt the crossed Amici prism design.

## 3. GLASS COMBINATIONS

The most critical issue facing the IRIS ADC design is whether there exist suitable glass pairs that can mimic the atmospheric dispersion sufficiently to meet the requirements. To answer this, we selected the NIR-transmitting glasses in Table 2, and then calculated best-fit linear combinations of glass pairs to the model atmospheric dispersion. This was performed over the entire wavelength range 0.8-2.4 μm to find those glass combinations with the best match to the overall atmospheric dispersion.

Table 2. Selected IR-transmitting glasses and crystals.

| High(er) Dispersion | Low(er) Dispersion |
|---|---|
| BAL42 | BAL42 |
| S-FTM16 | S-FTM16 |
| SiO2 | SiO2 |
| LiF | spinel |
| S-NPH2 | BaF2 |
| S-TIH11 | CaF2 |
| S-TIH14 | CSBR |
| S-TIH4 | N-PK51 |
| S-TIM28 | S-FPL51 |
| S-NPH1 | S-FPL52 |
| S-NPH53 | sapphire |
| S-LAH71 | BAL15Y |
| S-TIH53 | S-FPL53 |

The linear combination is described as $c_1(n_1-1)+c_2(n_2-1)$, where $n_1$ and $n_2$ are the refractive indices of the glasses. Prior to calculation, ZEMAX was used to adjust the refractive indices to the cryogenic environment of IRIS, that is, vacuum and T=77K. The atmospheric dispersion model was sampled on a constant-logarithm interval in wavelength (a linear interval produces comparable but slightly different results). A wavelength of zero-deviation near 1.4 μm was specified; again, choosing a different value produces similar but slightly different results. In the end, 163 unique viable glass combinations were examined. The best glass combinations are listed in Table 3, and the residual dispersions are shown graphically for the best four pairs in Figure 2a-d.

Table 3. Best glass combinations. Glasses marked (†) have poorer transmission at λ > 2.0 μm.

|   | Glass 1 (low ν) | Glass 2 (high ν) | rms (mas) | c1 ($10^{-4}$) | c2 ($10^{-4}$) | Ratio (lin. comb.) |
|---|---|---|---|---|---|---|
| 1 | S-NPH2 | Spinel | 2.65 | -1.664 | 2.069 | -0.804 |
| 2 | S-NPH1† | BAL42† | 3.12 | -2.573 | 3.472 | -0.741 |
| 3 | S-NPH1† | S-FTM16 | 3.28 | -4.543 | 6.101 | -0.745 |
| 4 | S-LAH71 | S-FPL52 | 3.77 | -2.825 | 5.149 | -0.549 |
| 5 | S-NPH2 | BAL42† | 5.18 | -1.744 | 2.660 | -0.656 |
| 6 | S-FTM16 | BAL42† | 4.70 | -7.961 | 8.001 | -0.995 |
| 7 | S-NPH1† | Spinel | 6.39 | -2.416 | 2.659 | -0.908 |
| 8 | S-NPH2 | S-FTM16 | 7.27 | -2.611 | 3.961 | -0.659 |

Two of the best combinations involve spinel (magnesium-aluminum oxide), which is now available in blanks produced by hot-pressed sintering. However, there is some concern that the optical scattering in this material will be too high to make it practical for our uses; this needs to be investigated. Most of the other combinations use either S-NPH1 and/or BAL42, both of which have significantly poorer transmission across the K-band compared to the other glasses.

The best pairs have rms residual dispersions of a few milliarcseconds (mas). The dispersion correction can be "tuned" to provide better correction within a single passband at the expense of the other passbands. In this case, the ratio of the coefficients is maintained while c1 and c2 are scaled by a constant amount to produce a bit-fit match to the dispersion with each passband. These results are shown in the lower panels in Figure 2a-d, with the scale on the left. We can see that we meet or come very close to our requirement (less than ±1 mas residual) in all passbands.

Finally, we explored how this residual dispersion might couple with spectral energy distribution (SED) to produce systematic errors in the astrometry. Five spectral types (O5V, A0V, G5III, M0III and M6III) were chosen from the stellar spectrophotometric atlas of Pickles[7] to span a typical range in stellar types. The residual dispersions within each passband were then weighted with the relative flux of the different SEDs and averaged. These values are also shown in the lower panels of Figure 2a-d as symbols, with the scale shown on the right. These systematic errors with spectral type are generally within ±20 microarcseconds, with the worst cases being the reddest stars in the Z and K passbands. Since this is at the level of the desired astrometric precision, it would appear that any of these glass pairs are suitable for the IRIS ADC.

As a baseline, we adopt the S-NPH2/spinel combination, but we obviously cannot commit to this pair until the potential scattering in spinel is investigated.

## 4. SPECIFIC OPTICAL DESIGN

In the last section, we identified glass pairs which satisfactorily mimic the atmospheric dispersion in linear combination. In practice, we must fabricate real prisms of these materials and second-order effects may become important. First, we must calculate the actual prism angles required in each glass pair. The large ratio of focal lengths of telescope and collimators means there is a large angular magnification in IRIS (~600× for the imager; ~4500× for the IFU), which applies to the required dispersion produced by the ADC. Table 4a lists typical prism angles and thicknesses needed for the imager; Table 4b lists them for the IFU. These values were calculated analytically, and then put into a ZEMAX model for verification. The ZEMAX models can also be used to explore second-order effects, such as off-axis behavior and distortion.

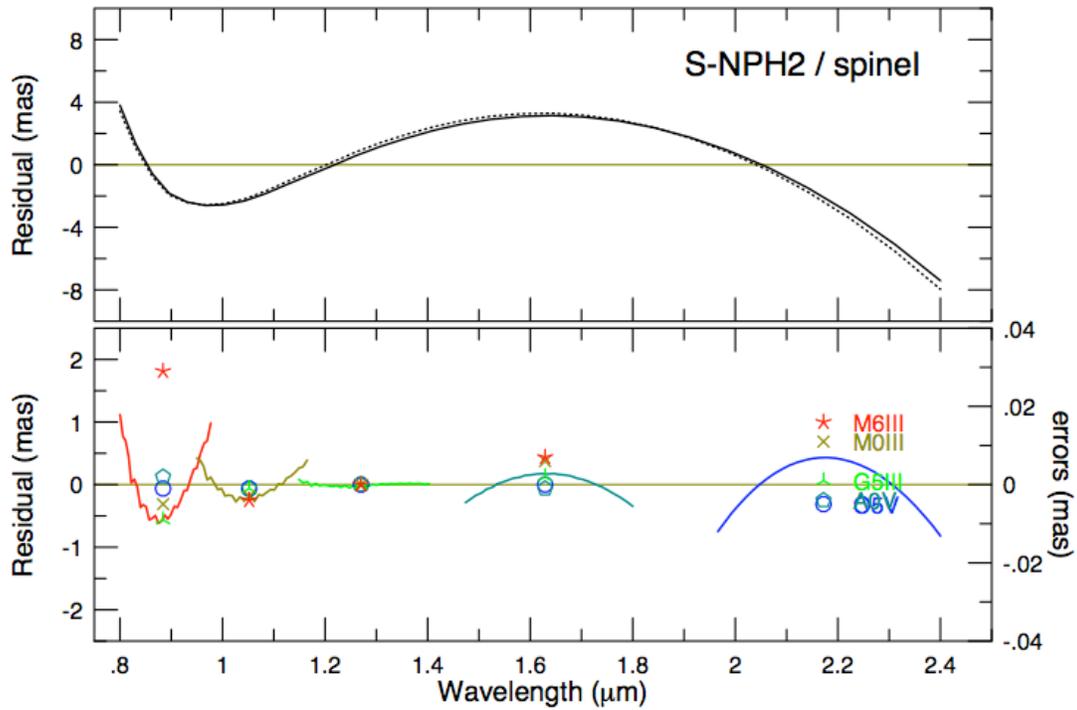

Figure 2a. The top panel shows the residual dispersion at Z=65° for the S-NPH2/spinel combination across the full wavelength range, 0.8-to-2.4μm, optimized for constant logarithmic (solid) and linear (dashed) wavelength intervals. The lower panel shows the same (logarithmic interval) combination "tuned" to minimize residuals in each passband. Such residuals couple with the SED of stars to produce astrometric errors, which are shown as symbols for 5 different SEDs (scale on the right). The five stellar SEDs are O5V, A0V, G5III, M0III and M6III.

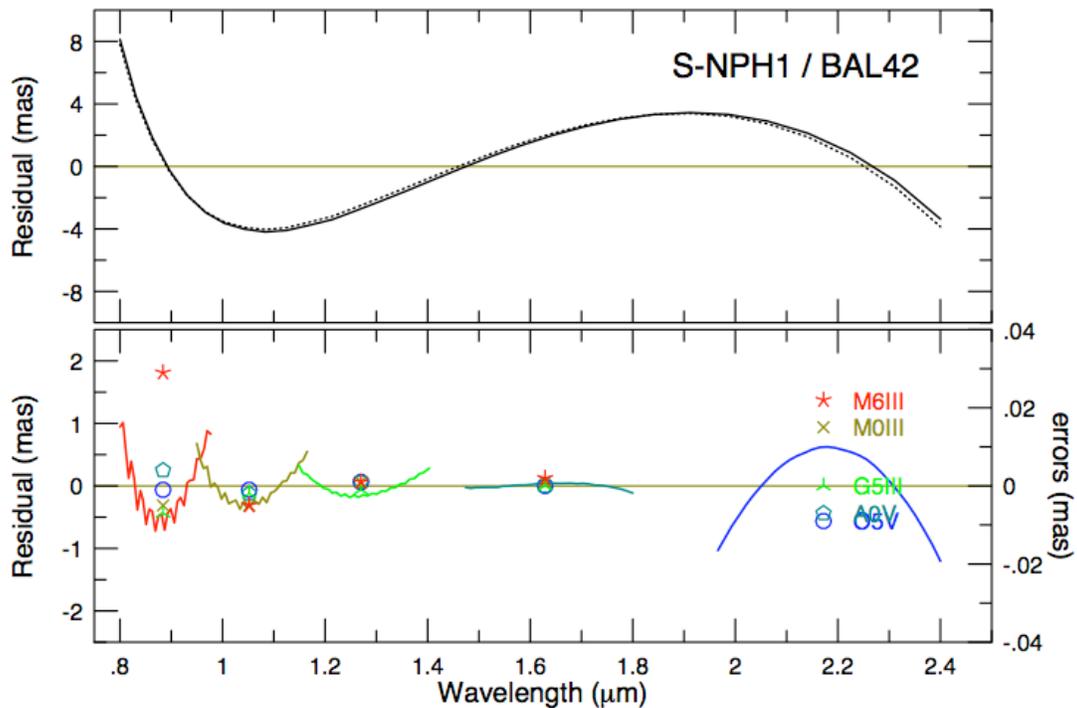

Figure 2b. (See caption above).

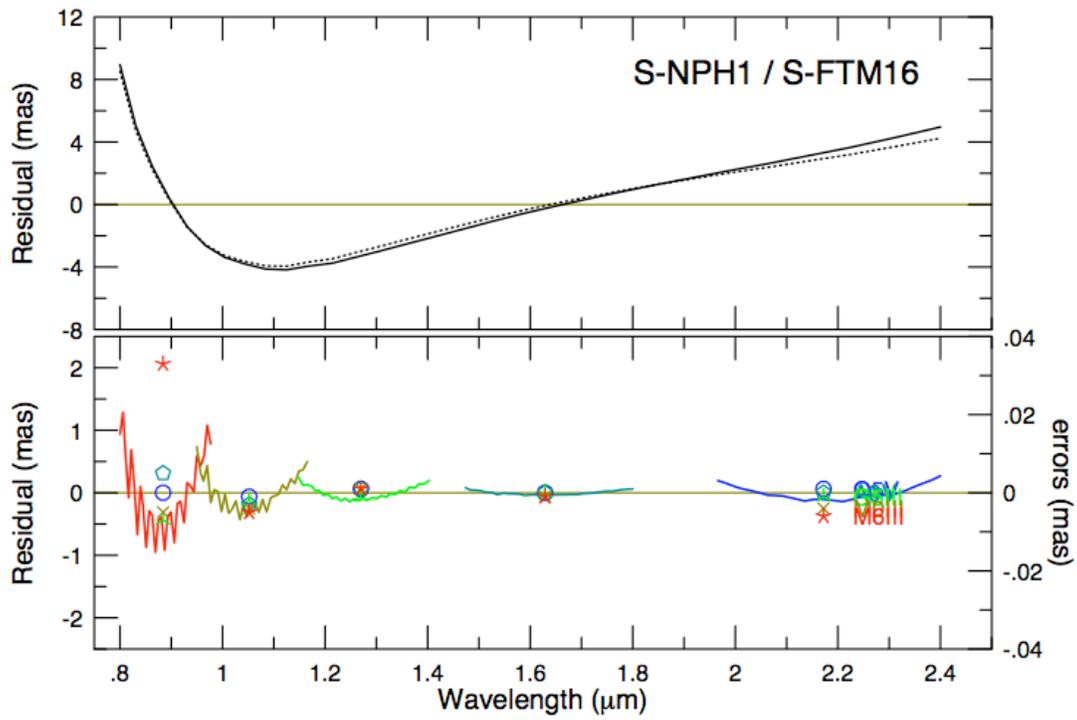

Figure 2c. (See caption above).

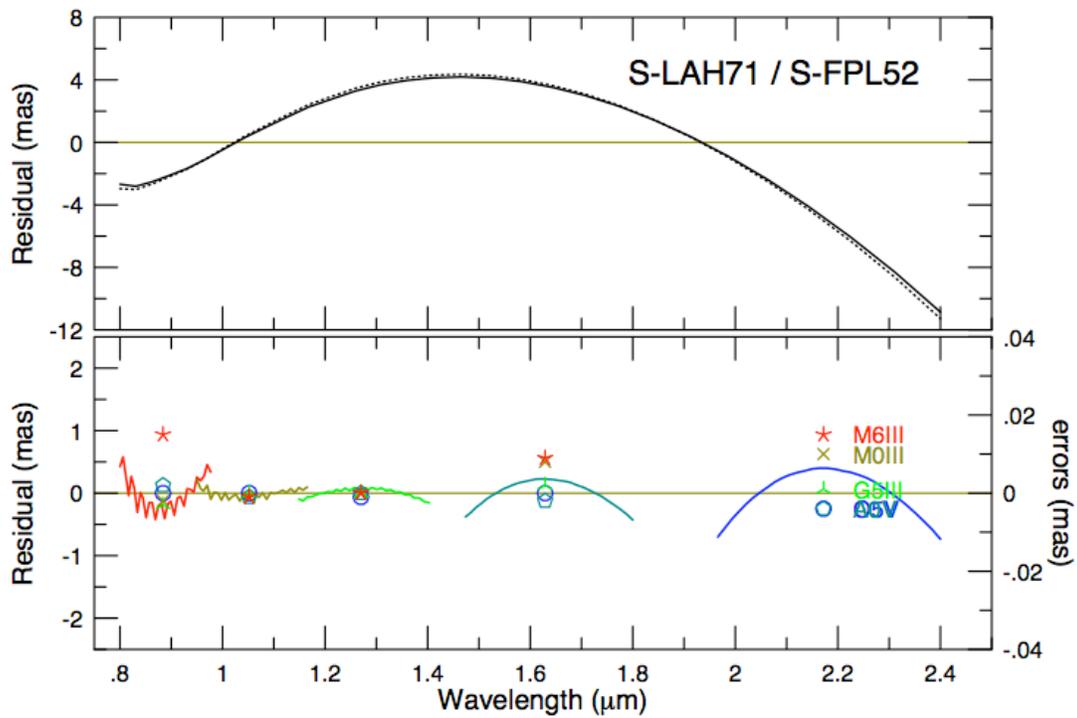

Figure 2d. (See caption above).

Table 4a. Prism parameters for the imager.

|   | Glass 1 (low ν) | Glass 2 (high ν) | Prism1 Angle (deg) | Prism2 Angle (deg) | Est. thick-ness (mm) |
|---|---|---|---|---|---|
| 1 | S-NPH2 | Spinel | 2.9194 | 3.6319 | 7.0, 7.5 |
| 2 | S-NPH1* | BAL42* | 4.5864 | 6.1967 | 8.2, 9.3 |
| 3 | S-NPH1* | S-FTM16 | 8.2029 | 11.0536 | 10.8, 12.8 |
| 4 | S-LAH71 | S-FPL52 | 4.8716 | 8.9178 | 8.4, 11.3 |
| 5 | S-NPH2 | BAL42* | 3.2278 | 4.9265 | 7.3, 8.4 |
| 8 | S-NPH2 | S-FTM16 | 6.1857 | 9.4170 | 9.3, 11.6 |

Table 4b. Prism parameters for the IFU.

|   | Glass 1 (low ν) | Glass 2 (high ν) | Prism1 Angle (deg) | Prism2 Angle (deg) | Est. thick-ness (mm) |
|---|---|---|---|---|---|
| 1 | S-NPH2 | Spinel | 18.2475 | 22.9636 | 9.1, 10.3 |
| 2 | S-NPH1* | BAL42* | 23.0790 | 32.1987 | 10.3, 12.9 |
| 3 | S-NPH1* | S-FTM16 | 26.3778 | 37.0484 | 11.2, 14.4 |
| 4 | S-LAH71 | S-FPL52 | 19.5308 | 38.9781 | 9.4, 15.1 |
| 5 | S-NPH2 | BAL42* | 18.3333 | 28.9512 | 9.1, 11.9 |
| 8 | S-NPH2 | S-FTM16 | 22.4944 | 36.1074 | 10.2, 14.1 |

Imager prism angles are modest (~3° to ~6° in most cases). However, given the relatively large FOV and high angular magnification, there is a concern that the dispersion on-axis will vary significantly from that at the edge of the field, where the incidence angle is not near normal. Therefore, we used the ZEMAX model to calculate the residual dispersion curve for points on-axis and near the top and bottom of the FOV. The results are shown in Figure 3. The residual dispersion curves are almost identical for the extreme field points, and they differ by a small but significant amount from the on-axis curve. The difference is <0.6 mas, but it is expected this difference will become insignificant within a given passband when the ADC is "tuned" for that passband.

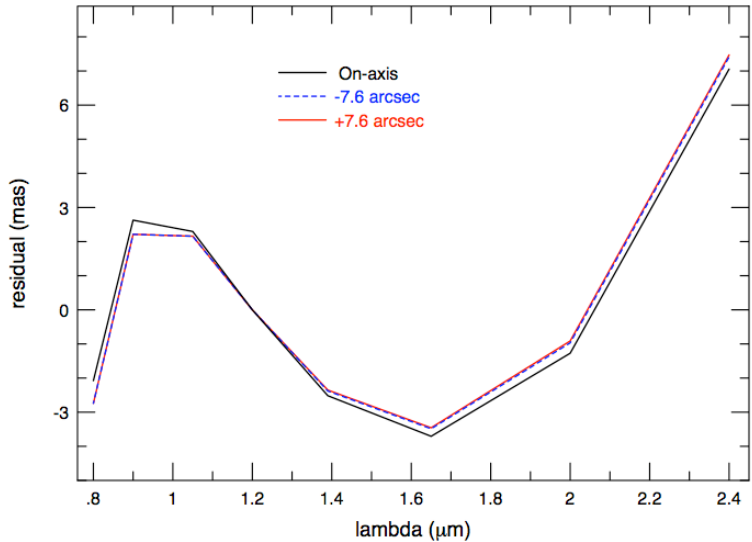

Figure 3. Residual dispersion of points at top, bottom and center of the imager field, as calculated by ZEMAX for the baseline model (S-NPH2/spinel). The difference in dispersion correction is small, and insignificant when the ADC is tuned to individual passbands.

The crossed-Amici prism design does not introduce image aberrations in a perfectly collimated beam, but is does introduce distortion. Figure 4 illustrates this problem. While the maximum distortion is under 1 mas, there are two obvious ramifications. First, for precision astrometry, this distortion will need to be calculated and removed. Secondly, a small amount of [field-dependent] image blur can occur as the elevation changes which could potentially limit exposure times.

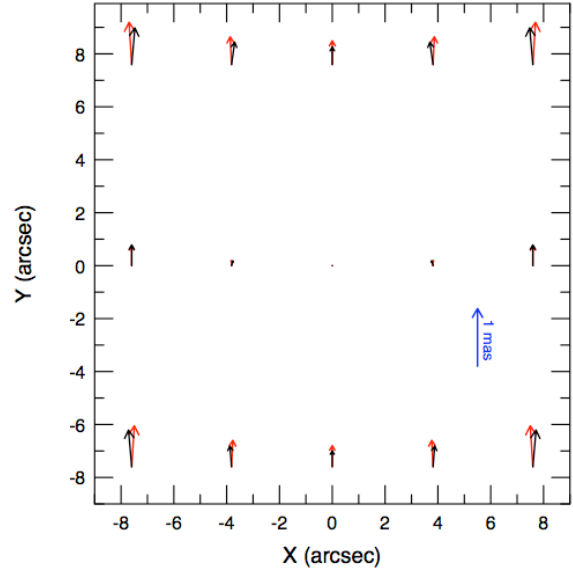

Figure 4. Elevation-dependent distortion at 1.2μm (black) and 2.4μm (red). The vectors represent the relative change in position (with respect to the on-axis image) between ZD=0° and 65°. This is the distortion introduced by the prisms alone, and does not include any distortion caused by differential atmospheric refraction.

## 5. SENSITIVITIES AND PRACTICAL LIMITATIONS

In this section, we wish to draw attention to some challenges and essential limitations caused by atmospheric refraction and dispersion on small scales, such as will be reached by IRIS/TMT. (See also the discussion by Helminiak[8].)

First, in order to correct for atmospheric dispersion, we need to know it precisely. How well can we predict the actual dispersion? For Mauna Kea, we adopted a reference case of model temperature 0°C, pressure 620 mbar, and 30% relative humidity. We then varied these three parameters to see how the refraction at 1.0 μm and zenith distance 60° was affected. Relative to the reference case, we found that refraction varied with pressure at a rate of $1.6 \times 10^{-3}$ / mbar, and with temperature at a rate of $-3.6 \times 10^{-3}$ / C. The change with relative humidity is negligible over its entire range. As an example of how well atmospheric conditions must be known, we note that a typical atmospheric dispersion across a passband is ~100 mas at a zenith distance of 60°. An error of 1 mas is of order $\sim 10^{-2}$ of the total dispersion (or differential refraction) across a passband, roughly equivalent to an error of 6 mbar in pressure or 3°C. Thus, at a minimum, atmospheric conditions will need to be carefully monitored and parameters updated in the dispersion model to achieve the required dispersion corrections, at least at lower elevations.

A second issue is differential atmospheric refraction across the field, which is a function of changing elevation across the field. This produces a "compression" of the apparent field along the parallactic angle. This effect is, of course, not removed by the ADC. As an example, at 65° zenith distance and 1.0 μm, this "compression" is about 10 mas across 10 arcsecond change in elevation; the compression is about 8 mas at 62°. Thus, across 3° in zenith distance objects at the top or bottom of the IRIS field could move 2 mas with respect to the center. This changing distortion is always aligned with the parallactic angle, so rotation of the field with respect to the parallactic angle will also produce image movement. Image blurring produced by these changing distortions can become significant for long exposures at lower elevations. In any event, these effects must be carefully modeled and accounted for in performing precision astrometry.

## 6. CONCLUSION

We have identified several glass combinations that are capable of correcting atmospheric dispersion over 1°≤ Z ≤65° to a level of ±1 milliarcsecond across the *Z*, *Y*, *J*, *H*, and *K* bands, thus meeting the requirements for IRIS. ZEMAX optical models confirm the performance, and also allow us to estimate the severity of field distortion, which appear manageable. We call attention to some challenges introduced by the atmosphere in terms of modeling the dispersion, and performing precision astrometry in the presence of differential refraction across the field of view of the imager.

The authors wish to acknowledge the support of the TMT partner institutions. They are the Association of Canadian Universities for Research in Astronomy (ACURA), the California Institute of Technology and the University of California. This work was supported as well by the Gordon and Betty Moore Foundation, the Canada Foundation for Innovation, the Ontario Ministry of Research and Innovation, the National Research Council of Canada, Natural Sciences and Engineering Research Council of Canada, the British Columbia Knowledge Development Fund, National Astronomical Observatory of Japan (NAOJ), the Association of Universities for Research in Astronomy (AURA) and the U.S. National Science Foundation.